\documentclass[twocolumn]{aastex63}
\usepackage{graphicx}

\usepackage{braket}
\usepackage{bm}
\usepackage{amssymb, amsmath}

\newcommand{\Msun}{\rm M_{\odot}}
\newcommand{\Ha}{H{$\rm \alpha$}~}
 
\newcommand{\OIII}{O{\sc iii}}

\submitjournal{ApJL}
\shorttitle{reconstruction of 3D intensity maps with ML}
\shortauthors{Moriwaki et al.}

\begin{document}

\title{Deep learning reconstruction of three-dimensional galaxy distributions
with intensity mapping observations}

\correspondingauthor{Kana Moriwaki}
\email{kana.moriwaki@phys.s.u-tokyo.ac.jp}

\author{Kana Moriwaki}
\affiliation{Department of Physics, The University of Tokyo, 7-3-1 Hongo, Bunkyo, Tokyo 113-0033, Japan}

\author{Naoki Yoshida}
\affiliation{Department of Physics, The University of Tokyo, 7-3-1 Hongo, Bunkyo, Tokyo 113-0033, Japan}
\affiliation{Kavli Institute for the Physics and Mathematics of the Universe (WPI),
UT Institutes for Advanced Study, The University of Tokyo, 5-1-5 Kashiwanoha, Kashiwa, Chiba 277-8583, Japan} 
\affiliation{Research Center for the Early Universe, School of Science, The University of Tokyo, 7-3-1 Hongo, Bunkyo, Tokyo 113-0033, Japan}
\affiliation{Institute for Physics of Intelligence, School of Science, The University of Tokyo, 7-3-1 Hongo, Bunkyo, Tokyo 113-0033, Japan}

\begin{abstract} 
 Line intensity mapping is emerging as a novel method that 
 can measure the collective intensity fluctuations of atomic/molecular line emission from distant galaxies.
 Several observational programs with various wavelengths are ongoing and planned,
 but there remains a critical problem of line confusion; 
 emission lines originating from galaxies at 
 different redshifts are confused at the same observed wavelength.
 We devise a generative adversarial network 
 that extracts designated emission line signals from noisy 
 three-dimensional data.
 Our novel network architecture allows two input data, in which the same underlying large-scale structure 
 is traced by two emission lines of \Ha and [\OIII], 
 so that the network learns the relative contributions at each wavelength and is trained to decompose the respective signals.
 After being trained with a large number of realistic mock catalogs, the network is able to 
 reconstruct the three-dimensional distribution of emission-line galaxies at $z = 1.3-2.4$.
 Bright galaxies are identified with a precision of 84\%, and the cross-correlation coefficients between the true and reconstructed intensity maps are as high as $0.8$.
 Our deep-learning method can be readily applied to data from planned space-borne and ground-based 
 experiments.
\end{abstract} 


\section{Introduction}
The large-scale distribution of galaxies carries rich 
information on the structure and the 
evolution of the Universe and on how galaxies are formed from early through to the present day.
Line intensity mapping (LIM) is aimed at measuring large-scale intensity fluctuations of line 
emissions from galaxies and intergalactic gas.
Complementary to traditional galaxy surveys, LIM covers a broad spectral range and detects signals
from essentially {\it all} emission sources residing in a large cosmological volume \citep{Kovetz17}. 
It is thus possible to make a structural "map"
of the Universe by a single observation.
There have already been a few successful experiments that detect hydrogen 21-cm line \citep{Chang10, Ali15}, and
observations targeting other emission lines such as CO/[C{\sc ii}] and $\rm Ly\alpha/H\alpha$/[\OIII] 
are ongoing \citep[e.g.,][]{Keating20, Concerto20, Cleary21} 
or planned \citep[e.g.,][]{Dore14, Dore18}. 
LIM can efficiently survey a large observational volume, 
and the data from LIM are well suited, for instance, to study
the formation and evolution of galaxies \citep{Breysse16, Keating16} 
as well as geometry and the matter content of the Universe
\citep[see, e.g.,][]{Dore14}.
LIM can also be used to study the reionization by combining with 21 cm observations of the inter-galactic medium \citep{Dumitru19, Moriwaki19}.

A key process in the analysis of LIM data is to separate the contributions from different emission lines originating from sources at different redshifts.
Let us consider two emission lines with rest-frame wavelengths $\lambda_1$ and $\lambda_2$.
If they are emitted at redshifts $z_1$ and $z_2$ that satisfy $\lambda_1(1+z_1) = \lambda_2(1+z_2) = \lambda_{\rm o}$,
they are observed at the same wavelength $\lambda_{\rm o}$, appearing as "interlopers" to each other.
Cross-correlation analyses are proposed to solve this line confusion problem \citep[e.g.,][]{Visbal10},
and there are several other statistical methods \citep[e.g.,][]{Gong14, Cheng16}.
It is technically challenging to isolate the contribution of a particular emission line and to infer the intensity distribution, but
a successful direct reconstruction of the three-dimensional distribution of the emission sources
would enhance the constraining power in cosmological studies as well as studies on the galaxy formation and evolution.
If the contamination of interloper lines can be removed, we are able to 
analyze the large-scale structure accurately \citep[see e.g.,][]{Fonseca17}
and also constrain the galaxy population by using methods such as the voxel intensity distribution \citep{Breysse17}.

Customized convolutional neural networks (CNNs) have been developed and applied to separate different emission line signals and to effectively de-noise a map \citep{Moriwaki20, Moriwaki21}, but such applications are limited to two-dimensional images 
without spectral information.
\citet{Cheng20} devise a reconstruction method that makes use of spectral analysis.
Their algorithm effectively extracts the source galaxies with multiple emission lines
brighter than a few times the noise level, 
but fainter signals still remain difficult to be detected.
In this {\it Letter}, we propose to utilize the spectral information in 
an efficient manner
so that a "machine" can learn the correlation of multiple emission lines at different wavelengths.
It is possible to perform a full three-dimensional reconstruction by using a LIM observation with a broad wavelength coverage.
This finally enables the reconstruction of the three-dimensional cosmic structure with LIM.

\section{Method}

We primarily consider NASA's SPHEREx\footnote{https://spherex.caltech.edu} mission to be launched in 2024
and identify the two brightest emission lines \Ha 6563 \AA~ and [\OIII] 5007 \AA~
as our target to be detected by SPHEREx.
We do not consider the other interlopers such as [O{\sc ii}] and H$\beta$. While the other lines' intensities are likely to be subdominant, they can also carry additional information in the spectral domain. Our method can easily be adjusted to deal with more than two emission lines, although the time needed for training may increase.

\subsection{Training data}

\begin{figure*}
\begin{center}
  \includegraphics[width=17cm]{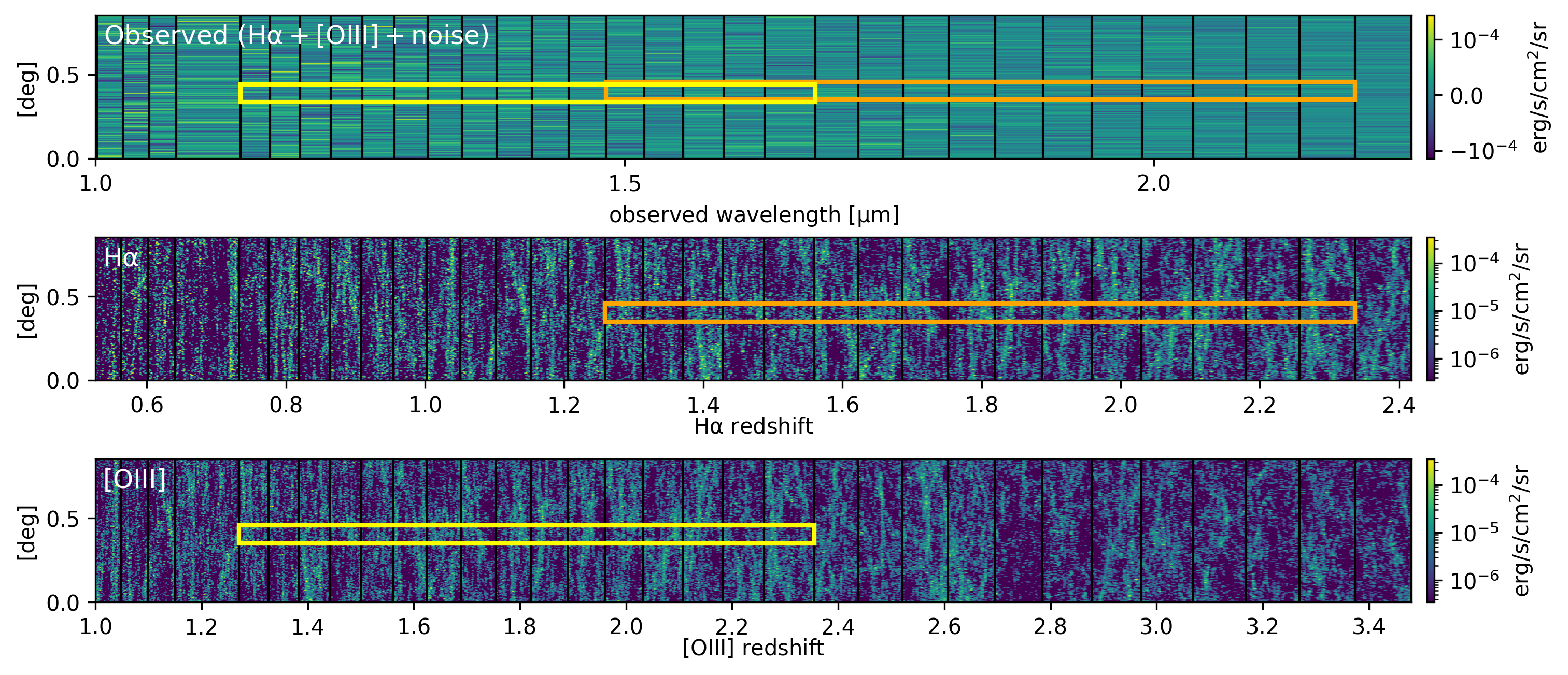}
  \caption{The intensity distribution on a past light-cone of a hypothetical observer having a 0.85 deg field-of-view. 
  The observed intensity (top), \Ha (middle) and [\OIII] (bottom) contributions in units of 
  erg $\rm s^{-1}~cm^{-2}~sr^{-1}$.
  The black lines show the spectral binning of the SPHEREx detector.
  The yellow and orange boxes indicate the redshift ranges of
  \Ha and [\OIII] emitters, whose signals
  originates from galaxies from $z=1.3$ to 2.4.
  The data cube within an angular size of $6.4'$ (the size of the yellow boxes) are used for training.
  Note that we have adopted larger angular resolution  for visibility in this figure than the actual resolution of our training data.
  }
\label{fig:lightcone}
\end{center}
\end{figure*}

\begin{figure}
\begin{center}
  \includegraphics[width=8cm]{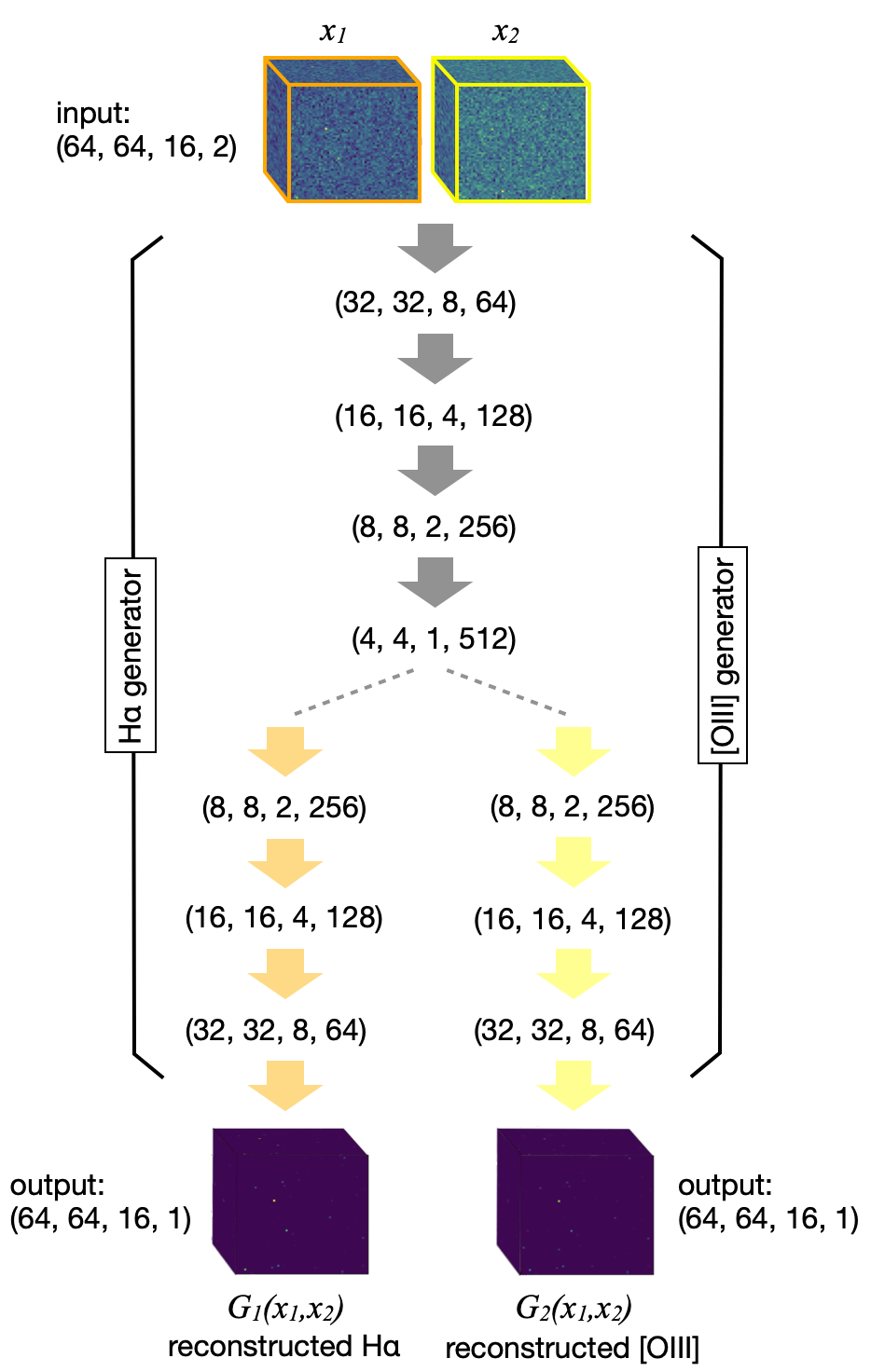}
  \caption{The architecture of the generator that takes two feature maps (data cubes) as an input and consists of four shared convolution layers, followed by four deconvolution layers. 
 }
\label{fig:gen-architecture}
\end{center}
\end{figure}

To generate mock observation catalogs for training and test, 
we use a publicly available code {\sc pinocchio} \citep{Monaco13}
that populates a large cosmological volume with dark matter
halos\footnote{We adopt $\Omega_{\rm m} = 0.316$, $\Omega_{\rm \Lambda} = 0.684$, and $h = 0.673$ \cite{Planck18}.}.
We configure past light-cones of a hypothetical observer 
by arranging several simulation outputs to fill the volume 
(Figure \ref{fig:lightcone}).
We set the simulation box size to $690 h^{-1}$ comoving Mpc and the aperture of the light cone to 1.5 deg.
The minimum halo mass considered is $2\times 10^{11}h^{-1}~\Msun$.
We have confirmed that the presence of smaller haloes does not affect the total intensity significantly nor the intensity distribution. 
We carefully choose the line-of-sight direction of the light cone so that any galaxy does not appear more than once in the redshift range of our interest.

To assign line luminosities to the galaxies (haloes), we use halo mass-to-line luminosity relations computed in our previous study \citep{Moriwaki20} based on the results of
cosmological hydrodynamics simulation IllustrisTNG \citep{Nelson19}.
We assign the luminosities by assuming that the line luminosities of haloes in a halo mass bin $M_i$ follow an asymmetric normal distribution 
with different variances on the larger and smaller side than the most frequent luminosity value $L_i$.
This assigning process produces similar scatter in the halo mass-to-line luminosity relations as that of IllustrisTNG.
Both the \Ha and [\OIII] line luminosities are approximately proportional to the star formation rate, 
but the derived \Ha/[\OIII] ratio varies over a factor of ten
because the [\OIII] luminosity depends also on the properties of the interstellar medium such as metallicity and ionization parameter.
We find that the line ratios of our catalog haloes are also scattered in a similar way as that computed with IllustrisTNG.

The middle and bottom panels of Figure \ref{fig:lightcone} show the intensity distributions of \Ha and [\OIII] on a past light-cone.
We adopt the spatial and spectral resolutions and the noise levels of SPHEREx \footnote{We use data in https://github.com/SPHEREx/Public-products}.
The angular resolution is 0.1 arcmin and the spectral resolution is approximately constant ($R \sim 40$) over the wavelength range of our interest.%
\footnote{The spectral resolution (binning) is not always constant. For example, there are wider bins at around
$1.1~\rm \mu m$. We do not use such irregular bins.}  
Note that the corresponding physical length to the angular resolution is much smaller than that of the spectral resolution.
At $z = 1.5$, for instance, 
$0.1$ arcmin corresponds to 52 kpc, 
while $R \sim 40$ corresponds to 47.2 Mpc. 
We add Gaussian noise to make realistic mock catalogs.
The noise level is about two orders of magnitude larger than the mean intensities of line emissions (top panel of Figure \ref{fig:lightcone}),
and thus detecting diffuse sources that are distributed over the entire intensity field is difficult even with our machine learning method.

For training, we generate 500 independent light-cones with 1.5 deg aperture over $\lambda_{\rm obs}= 1.0\rm \mu m-2.5~\rm \mu m$ using {\sc pinocchio}.
The wavelength range corresponds to 32 spectral bins of SPHEREx as shown in Figure 1.
To reduce computational cost, we generate input data with $64 \times 64$ angular pixels.
This corresponds to a field-of-view of $6.4' \times 6.4'$ with an angular resolution of 0.1 arcmin.
From each light cone, we randomly extract 100 
such small volumes. Then a total of 50,000 mock observational data cubes are
generated.
As discussed in the following section, we use two portions of the mock observational data with different wavelength ranges (indicated by orange and yellow boxes in Figure 1) 
as input to the neural networks.

\subsection{Network}

We use a conditional generative adversarial network \citep[cGAN;][]{Isola16} to perform 
the three-dimensional reconstruction.
In particular, we adopt conditional Wasserstein GAN \citep[WGAN;][]{Arjovsky17}.
WGAN is known to increase training stability and the diversity of generated data \citep{Foster19}. 
We have four 3D convolutional neural networks: 
two generators, $G_1$ and $G_2$, that reconstruct \Ha and [\OIII] signals from observed data 
and corresponding two critics
\footnote{In WGAN, a network that works as a discriminator in vanilla GANs is called critic.}
, $D_1$ and $D_2$, 
that distinguish true and reconstructed images.
Each generator consists of four convolution layers followed by four de-convolution layers
(Figure \ref{fig:gen-architecture}), whereas
the critic consists of four de-convolution layers.
The networks also include skip connections \citep{Isola16}, dropout \citep{Srivastava14}, and batch normalization \citep{Ioffe15}. 

The most important information to be learned by the generators is the co-existence of multiple emission lines at different wavelengths.
To make it easier for the generators to learn that the two emission lines are always observed with a separation of
$\Delta \lambda_{\rm obs} = (\lambda_{\rm H\alpha} - \lambda_{\rm [OIII]}) \times (1+z)$, 
we arrange the architecture so that the generators receive a pair of observed data cubes as an input.
The cubes are covered by sixteen SPHEREx wavelengths filters from $1.48~\rm \mu m$ to $2.19~\rm \mu m$, and $1.14~\rm \mu m$ to $1.68~\rm \mu m$,
which correspond to $1.25 \lesssim z \lesssim 2.4$ of \Ha and [\OIII] lines, respectively.
The input cubes, denoted by $x_1$ and $x_2$, are indicated by the orange and yellow boxes in Figure \ref{fig:lightcone}.
By giving the two data cubes arranged such that the two emission lines from the same source appear at the same pixel, 
we let the generators learn the consistent co-existence of the two lines. 

The critics also receive two data cubes as an input:
either a pair of the observed and reconstructed data, $(x_i, G_i(x_1,x_2))$,
or a pair of observed and true data, $(x_i, y_i)$, 
where $y_i$ is the true data cubes of \Ha ($i=1$) or [\OIII] ($i=2$) that cover the same wavelength range as $x_i$.

The networks are trained to optimize two loss functions defined by
\begin{align}
	L_i = D_i(x_i, y_i) - D_i(x_i, G_i(x_1, x_2)) + \lambda_i |y_i-G_i(x_1, x_2)|,
\end{align}
where the indices $i = 1, 2$ correspond to \Ha and [\OIII],
and $\lambda_i$ is a hyperparameter which we set $\lambda_1 = \lambda_2 = 100$ after some experiments.
The objective of the generators (critics) is to decrease (increase) the loss functions.
Another important building block of WGAN is the Lipschitz constraint imposed on the critics,
which prevents the outputs of the critics from changing abruptly.
To enforce the constraint, 
we adopt the same approach as in the original proposal by \citet{Arjovsky17} 
in which they clip the weights of the critic to lie within a small range of $[-0.01, 0.01]$.

We build our network using Tensorflow.
We use Adam optimizer \citep{Kingma14} with a learning rate of 0.0002 for training, set the batch size to be 50, and run 50 epochs on a single Nvidia Titan RTX GPU.

\section{Result}

To measure the performance of our WGAN, we generate an additional set of 1000 light-cones that are independent of the training data.
We randomly choose an area of $0.85~\rm deg \times 0.85~\rm deg$ from each light-cone
and divide it into $8\times 8$ cubes with the same size as the training data. 
The prepared test data are given to the generator of our WGAN. 
Finally, we reconstruct intensity cubes by combining $8\times 8$ outputs.

\begin{figure}
\begin{center}
 \includegraphics[width=8cm]{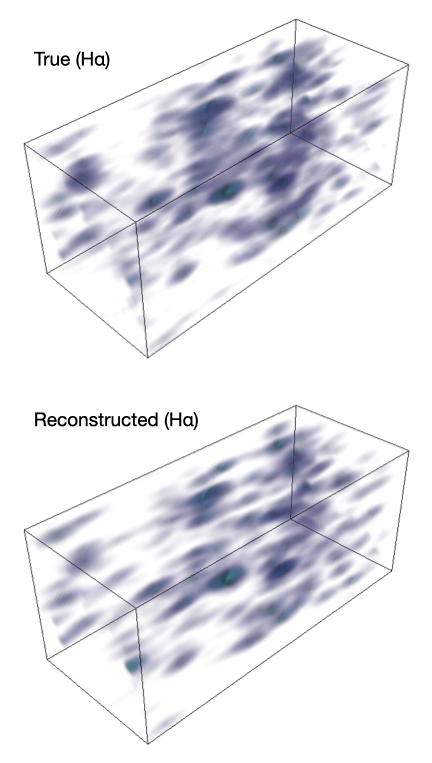}
 \caption{The true (top) and reconstructed (bottom) intensities of \Ha line emission from $z = 1.3$ to 2.4. The angular size is $\rm 0.43~deg \times 0.43~deg$. The intensities are smoothed for visibility with 6 and 0.5 times the pixel size for
 angular and spectral domain, respectively.}
\label{fig:true_rec_ha}
\end{center}
\end{figure}

\begin{figure}
\begin{center}
 \includegraphics[width=6cm]{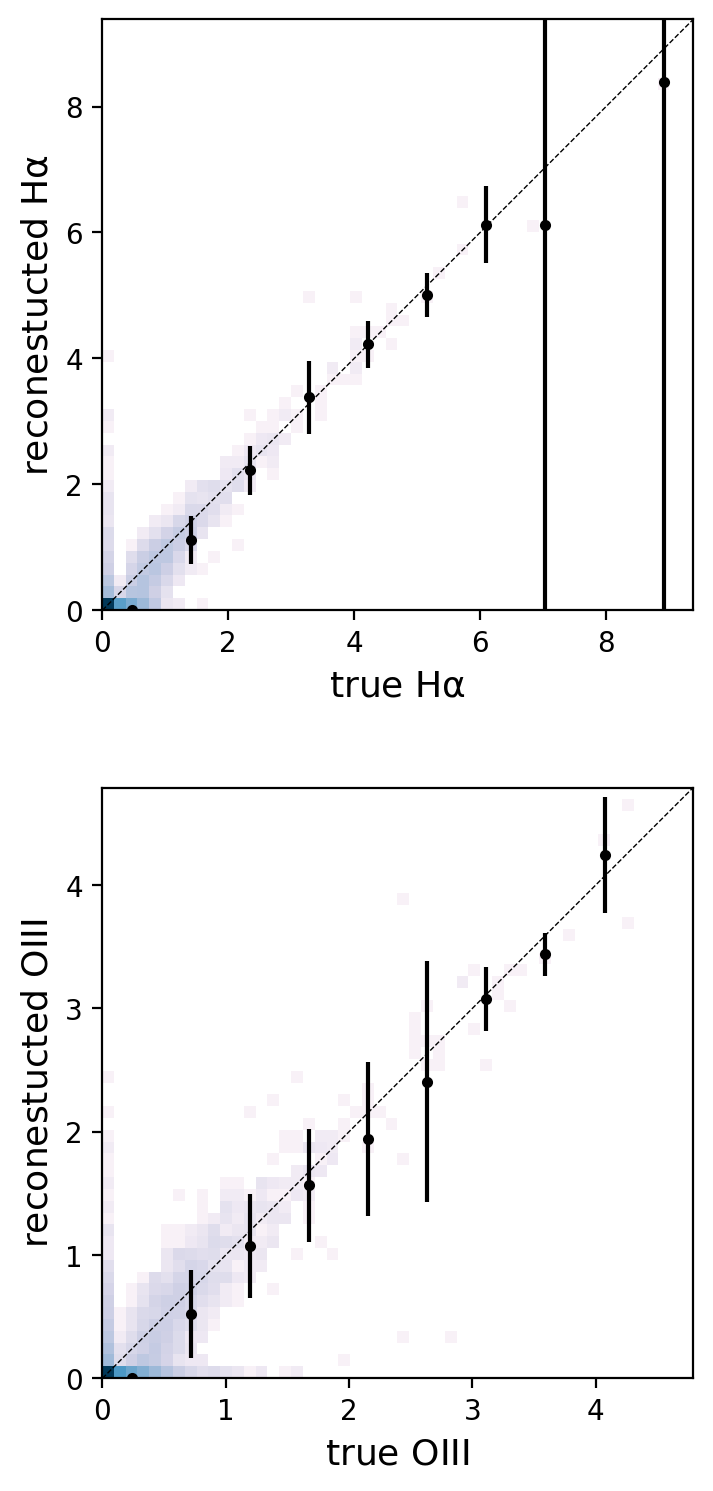}
 \caption{Pixel-by-pixel correspondence between the true and reconstructed intensities of \Ha (top) and [\OIII] (bottom). 
 Intensities are normalized by $10^{-5}~\rm erg/s/cm^2/sr$.}
\label{fig:pix-pix}
\end{center}
\end{figure}

\begin{figure}
\begin{center}
  \includegraphics[width=8cm]{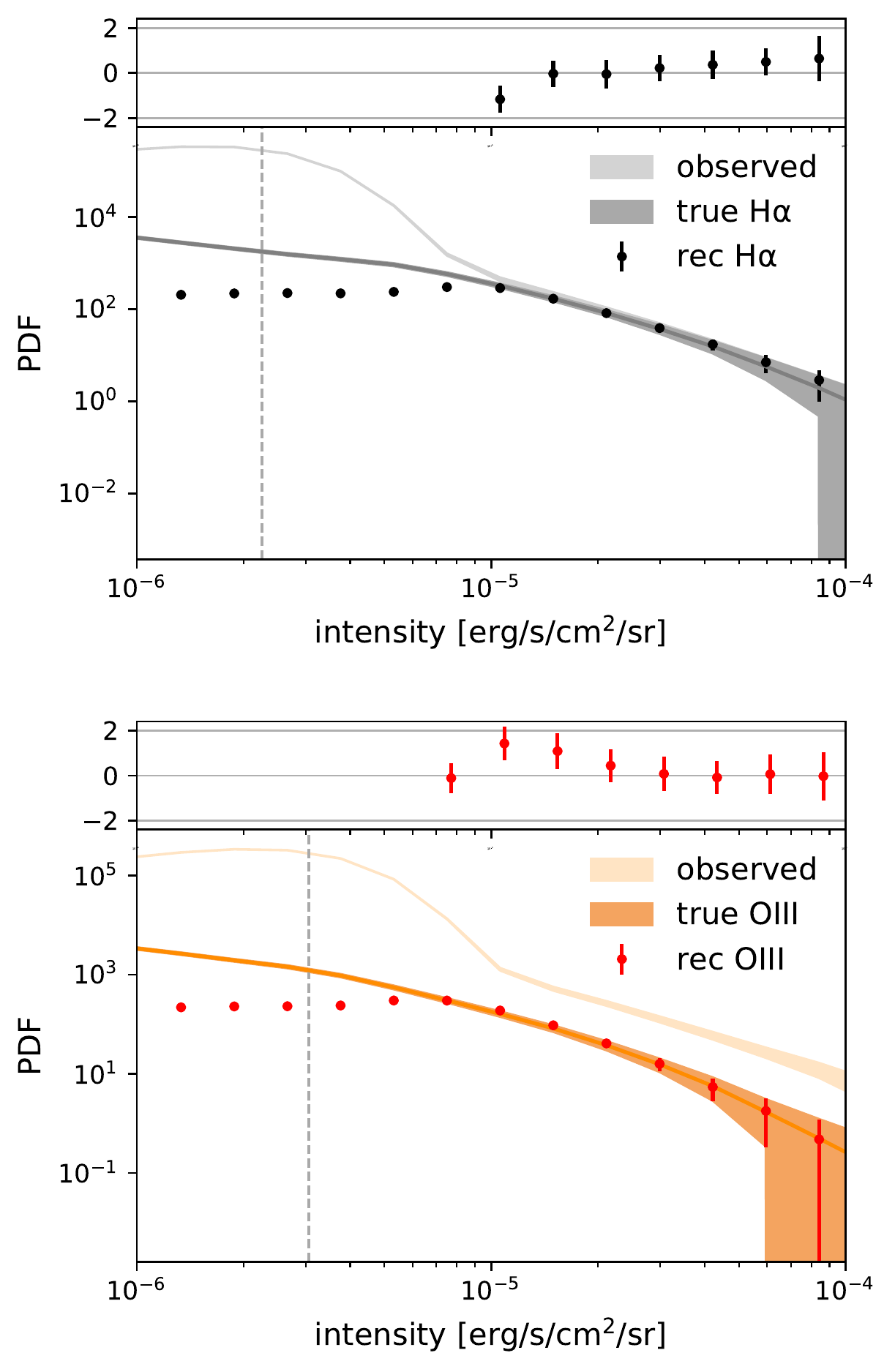}
  \caption{The one-point distribution function (PDF) of the true and reconstructed maps of \Ha (left) and [\OIII] (right). 
  The 1-$\sigma$ variation of the observed (light shades), true (dark shades) and reconstructed (error bars) PDFs over 1000 test data are shown.
 The dashed vertical lines are the noise level of SPHEREx, $\sigma_n$, averaged over 16 wavelength bins of the input data cubes.}
    \label{fig:pdf}
\end{center}
\end{figure}

In Figure \ref{fig:true_rec_ha}, we show an example of the true and reconstructed \Ha intensity distributions from $z = 1.3$ to 2.4.
The large-scale galaxy distribution is reproduced accurately
in 3D, despite the large noise level (see Figure \ref{fig:lightcone}).
Pixel-by-pixel comparison shows remarkably good agreement between the true and reconstructed maps (Figure \ref{fig:pix-pix}).
Our network reconstructs the brightest sources accurately, and thus the underlying large-scale distribution is also well reproduced.
Diffuse sources are not well reproduced 
because of the large observational noise considered in our study.
This can also be seen in the point distribution function (Figure \ref{fig:pdf}).
The bright ends are reproduced, but the WGAN appears to have learned that it is optimal to regard faint pixels just as noise-dominated.
The vertical lines are the noise level of SPHEREx averaged over 16 wavelength bins of the input data cubes, 
$\sigma_n = 2.25\times 10^{-6}$ (upper), $3.06\times 10^{-6}~\rm erg/s/cm^2/sr$ (bottom).
Figure 5 indicates that the effective limit of our machine learning reconstruction is a 
few-$\sigma_{\rm n}$. This is similar to the result of \citet{Cheng20}, who show that the CO line signals from similar redshifts are reconstructed down to a few-$\sigma_{\rm n}$  level.
Detecting diffuse "clouds" would be extremely difficult unless the observational noise is significantly reduced in future experiments.
It should be noted here that the weaker [\OIII] signals are also accurately reconstructed, 
even though the bright end of the observed PDF is dominated by foreground \Ha intensities.

We count the numbers of the pixels with intensities larger than 3-$\sigma_n$
in true ($N_{\rm true}$) and reconstructed ($N_{\rm rec}$) maps.
We then compute the recall, $N_{\rm X}/N_{\rm true}$, 
and the precision, $N_{\rm X}/N_{\rm rec}$, 
where $N_{\rm X}$ is the number of pixels that are detected
and matched in both the true and reconstructed maps.
The recall and the precision are 0.67 and 0.84 for H$\alpha$,
and the corresponding values for [\OIII] are 0.78 and 0.68.
We estimate that the typical intensities of [\OIII] are roughly half of H$\alpha$ at the same observed wavelength,
and our previous study shows that the detection performance degrades for such weaker lines
when only two-dimensional data is used for the machine learning analysis \citep{Moriwaki20}.
The impressive reproducibility of the [\OIII] distribution in the present study can be attributed to the inclusion of the spectral information, as we discuss in the following.

To quantify the reconstruction accuracy of the large-scale distribution, 
we compute the cross-correlation coefficient
\begin{align}
	r(k) = \frac{P_{X}(k)}{\sqrt{P_{\rm true}(k)P_{\rm rec}(k)}},
\end{align}
where $P_{X}$ is the cross-power spectrum and $P_{\rm true}$ and $P_{\rm rec}$ are the auto-power spectra of the true and reconstructed maps.
We find that a high reconstruction performance with $ r \sim 0.8$ at $k = 0.3~\rm arcmin^{-1}$ 
for both \Ha and [\OIII] has been achieved over the wide redshift range.
This is consistent with the point source detection accuracy discussed above.


The high reproducibility of weaker [\OIII] signals suggests 
that the [\OIII] generator refers to the \Ha intensities
that are more easily reconstructed from the two inputs.
This is exactly what we expect the machine to learn, and it is important to understand how much 
it depends on the \Ha intensity.
To investigate the learning process further, we generate test data 
with different, uncorrelated realizations for \Ha and [\OIII] and feed to the [\OIII] generator. 
The result shows that the reconstructed [\OIII] map is biased toward the 
true \Ha map, indicating that the [\OIII] generator strongly relies on the input $x_1$
that includes the \Ha signals rather than the input $x_2$.
However, the test case yields the cross-correlation coefficients between the reconstructed \Ha and [\OIII] maps
that are smaller than the real case with actual \Ha - [\OIII] pairs by 0.2.
This indicates that the information on the weak [\OIII] line in the observed maps is still used to reconstruct accurately [\OIII] intensity distributions.

To examine if the spatial clustering information is used along with the spectral information,
we perform an additional test.
We randomly shuffle the pixels of the test data 
and get rid of the angular correlation in the signals while preserving the spectral correlation. We then input the shuffled data into our network. The test result shows that the network still achieves high reproducibility;
the bright pixels ($>10^{-5}~\rm erg/s/cm^2/sr$) are reproduced with similar precision of $\sim 0.6 - 0.8$ for both the lines.
This implies that our network emphasizes the spectral information 
(emission line features) more
than the spatial correlation information.
We note that we consider a small area of 
$6.4 \times 6.4 ~{\rm arcmin}^2$ for the reconstruction in this study.
With the finest resolution achievable for our available computational resources, we are able to represent point sources but the particular configuration does not allow incorporating 
large-scale clustering features.
In our previous study \citep{Moriwaki21}, 
we showed that the information on the large-scale clustering is 
more properly used when the training data are generated with a sufficiently large area.
Clearly, there is room for improvement in our method.
In our future work, we will use data set with larger dimensions 
so that a machine can learn both the spectral information
and the large-scale clustering of galaxies.

\section{Summary}
We have developed, for the first time, neural networks
that extract signals of two emission lines from noisy data obtained in LIM observations.
Our 3D WGAN makes use of the information on the co-existence of two emission lines
in a given pair of data cubes.
It is able to reconstruct the bright sources when trained with a large number of mock observational maps that are closely configured for the SPHEREx experiment.
Our method can be extended and applied to LIM observations at any other wavelengths.
Once we can extract the individual signals, the reconstructed data can be used for cosmological/astrophysical parameter estimate, cross-correlation analysis, and planning follow-up observations.

\section*{Acknowledgements}
We thank the anonymous referee for helpful suggestions and constructive remarks on our manuscript. 
We thank Masato Shirasaki for helping to develop the networks.
KM is supported by JSPS KAKENHI Grant Number 19J21379
and by JSR Fellowship.
NY acknowledges financial support from JST AIP Acceleration Research Grant Number JP20317829.

\bibliography{bibtex_library} 

\label{lastpage}

\end{document}